\documentclass[aps,prl,twocolumn,superscriptaddress]{revtex4}
\usepackage{graphicx}
\usepackage{morefloats}
\usepackage{color}
\usepackage{amssymb}
\usepackage{float}
\usepackage{amsmath}  
\usepackage{esvect}
\usepackage{wasysym}

\begin{document}

\newcommand{\ham}{\mathcal{H}} 
\newcommand{\Q}{\mathcal{Q}} 

\title{Exploring magnetic resonance with a compass} 

\author{Esther Cookson}
\affiliation{Physics and Astronomy, University of California, Riverside, CA 92521, USA}
\author{David Nelson}
\affiliation{Physics and Astronomy, University of California, Riverside, CA 92521, USA}
\author{Michael Anderson}
\affiliation{Physics and Astronomy, University of California, Riverside, CA 92521, USA}
\affiliation{Physics Teacher Academy, University of California, Riverside, CA 92521, USA}
\author{Daniel L. McKinney}
\affiliation{Santa Rosa Academy, Menifee, CA 92584, USA}
\affiliation{Physics Teacher Academy, University of California, Riverside, CA 92521, USA}
\author{Igor Barsukov}
\email[]{igorb@ucr.edu}
\affiliation{Physics and Astronomy, University of California, Riverside, CA 92521, USA}
\affiliation{Physics Teacher Academy, University of California, Riverside, CA 92521, USA}


\begin{abstract}
Magnetic resonance plays an important role in today's science, engineering, and medical diagnostics. Learning and teaching magnetic resonance is challenging since it requires advanced knowledge of condensed matter physics and quantum mechanics. Driven by the need to popularize this technologically impactful phenomenon, we develop an inexpensive table-top demonstration experiment. It unveils the magnetic resonance of a hand-held compass in the magnetic fields of a permanent magnet. The setup provides an immediate visualization of the underlying physical concepts and allows for their translation to broad student audiences.
\end{abstract}

\maketitle

\section{1. Introduction}
Magnetic resonance plays an essential role in today's science and technology. Magnetic resonance imaging is an indispensable noninvasive tool in medical diagnostics and research. Nuclear magnetic resonance spectroscopy is regularly used in physics, chemistry, biology, and materials sciences for investigating physical and chemical properties and fingerprinting of substances. Electron spin resonance is used for the detection and identification of free radicals and for controlling electron spin qubits in the quantum-computation. Electron spin resonance in magnetic materials allows for studying magnetic properties and controlling magnetic states in spintronics applications.

Magnetic resonance is a quantum phenomenon and describes a resonant interaction of spins with electromagnetic fields. Its understanding requires advanced knowledge of quantum mechanics and condensed matter physics and thus often remains elusive to students. However, the great impact of this phenomenon on medicine, science, and technology makes magnetic resonance an important subject for general education, and it requires a graspable model for translating the physical concepts without resorting to advanced physics.

One of the most common and familiar objects related to magnetism is the compass. Here, we develop  an inexpensive table-top demonstration experiment for magnetic resonance. The compass is placed into the magnetic field of a permanent magnet (large fridge magnet); an alternating field of a electromagnetic coil (solenoid) excites a resonant oscillation of the compass needle. The experiment has been proven to catch attention of audiences of all ages. The physical concepts of magnetic resonance are translatable via the demonstration experiment to students of various preparedness levels. The prerequisites are basic understanding of magnetism and the concept of resonance.  

Science students often lack the connections between concepts and physical situations. This demonstration is designed to provide this connection by allowing students to observe a result that is predicted by mathematical calculation. This demonstration experiment can be used at the high-school as well as the university level. High-school instructors should reduce the mathematical component of the experiment to a conceptual level. The experiment is further aiming at improving the scientific literacy \cite{Goodstein1992}.

\section{2. Description of the experiment}
A compass is placed on a desk, its needle aligns with the Earth magnetic field lines and points towards the Earth magnetic poles (Fig.\,1(a)). When a permanent magnet is brought close to the compass, the needle realigns towards the permanent magnet (Fig.\,1(b)); the magnetic field in the vicinity of a permanent magnet is much larger than the Earth field.

The compass can also be realigned by magnetic field produced by an electromagnetic coil. The magnet is placed in the middle of the coil, and a large constant current is supplied through the coil. The compass needle aligns with the coil's magnetic field, perpendicular to the coil plane (Fig.\,1(c)). The needle reverses its direction by 180\,deg when the polarity of the current is switched.

When instead of the constant current, a weaker alternating current is supplied to the coil, the compass can no longer follow the alternating magnetic field and presents a barely visible jitter. With a permanent magnet brought close to the compass as shown in Fig.\,1(d), the needle aligns towards the permanent magnet. The jitter remains very small with the permanent magnet being close to the compass or far away.

Now, the permanent magnet is first brought very close to the compass and then moved away slowly. At a certain intermediate distance between the compass and the magnet, the jitter of the needle increases strongly. The needle starts to notably oscillate. When the magnet is pulled farther away from the compass, this oscillation fades away. By moving the magnet back and forth, it is possible to identify the "sweet spot" where the oscillation of the compass needle occurs.

\begin{figure*}[t]
\centering
\includegraphics[width=0.7\textwidth]{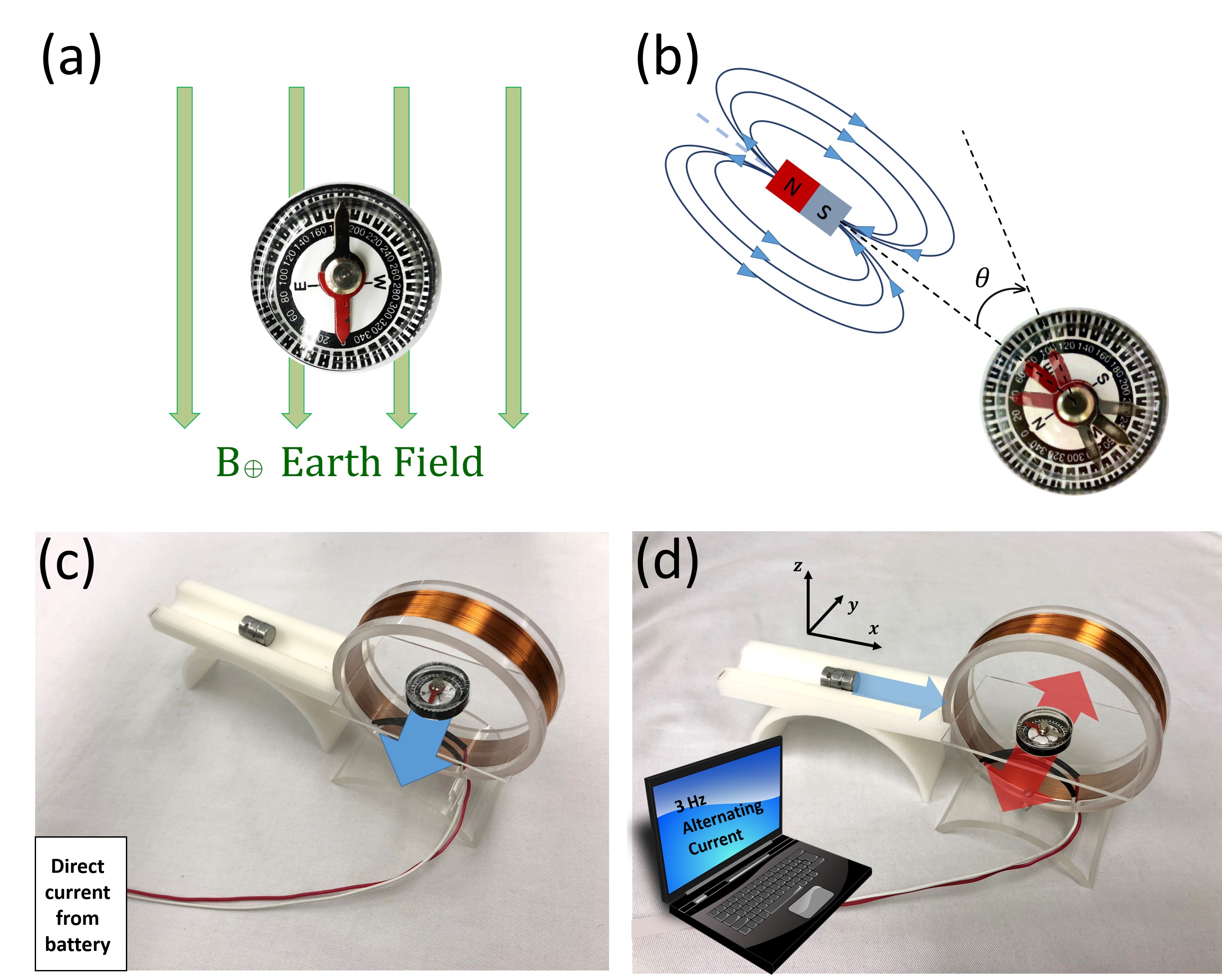}
\caption{(a)~A compass in Earth's magnetic field. (b)~The compass in close proximity to a permanent magnet, the needle points towards the magnet. When resonantly driven, the needle oscillates with the angle $\theta$.  (c)~Compass in a strong constant field generated by electromagnetic coil. (d)~Compass in a constant field of the permanent magnet and a weak alternating magnetic field of the coil. The fields are mutually perpendicular.}
\end{figure*}

\section{3. Theoretical concept}
The needle of the compass consists of a ferromagnetic material with magnetization $M_{\mathrm{needle}}$. The needle remains magnetized along its axis and posses a magnetic moment $\vec{m} = M_{\mathrm{needle}} V \vec{e}_{_{\mathrm{NS}}}$, where $V$ is the volume of the needle and $\vec{e}_{_{\mathrm{NS}}}$ is the unit vector pointing from south pole to north pole of the needle.

In the magnetic field \footnote{Strictly speaking, $\vec{B}$ is magnetic flux density} of the permanent magnet $\vec{B}_{\mathrm{PM}}$, the needle experience a torque 
\begin{eqnarray}
\label{torques}
\vec{T}=\vec{m}\times\vec{B}_{\mathrm{PM}}\, .
\end{eqnarray}
The torque tends to align the needle parallel to the external field. The field in the vicinity of a typical permanent dipole magnet is of the order of ${\sim5} \times 10^{-3}$ Tesla, which is much larger than the Earth magnetic field ($B_{\oplus}\approx 50 \times 10^{-6}$ Tesla). The compass serves as a simplistic detector of magnetic field.

The field in the center of a short electromagnetic coil is $\vec{B}_{\mathrm{coil}}\approx \mu_0 \frac{N I}{D}\,\vec{e}_y$; where $\mu_0=4\pi\,10^{-7}$\,Tesla$\cdot$m/Amp\`{e}re is the magnetic permeability of free space, $N$ is the number of windings, $D$ is the diameter of the coil, and $I$ is the electrical current through the coil. The direction of the field is perpendicular to the coil's plane, i.e. parallel to the $y$-axis (Fig.\,1(d)).

A constant current can produce a field larger than Earth field $B_{\oplus}$ and thus realign the compass. An alternating current produces an alternating drive field $\vec{B}_{\mathrm{drive}}(t)$. Generally, the compass cannot fully follow the alternating drive field due to inertia, instead it exhibits a slight jitter.

The compass lies in the horizontal plane in the center of the coil. The permanent magnet is aligned along the $x$-axis. The torque due to the field of the permanent magnet and the coil are vectorial entities, oriented along the $z$-axis. The sum of the torques is equal to the time-derivative of the angular momentum of the needle (Newton's axiom),
\begin{eqnarray}
\label{torques}
\vec{m}\times\vec{B}_{\mathrm{PM}} + \vec{m}\times\vec{B}_{\mathrm{drive}} =  \ddot{\theta} J \vec{e}_z\, ,
\end{eqnarray}
where $J$ is the moment of inertia and $\ddot{\theta}$ is the second time-derivative of the deflection angle of the needle. The $z$-component of Eq.\,\ref{torques} reads
\begin{eqnarray}
\label{torquesscalar}
m B_{\mathrm{PM}}\sin(-\theta) + mB_{\mathrm{drive}}\sin(\frac{\pi}{2}-\theta) =  \ddot{\theta} J \, .
\end{eqnarray}

In the small angle approximation: $\sin(-\theta)\approx-\theta$ and $\sin(\frac{\pi}{2}-\theta)\approx 1$, and Eq.\,\ref{torquesscalar} becomes:
\begin{eqnarray}
\label{SHO}
\ddot{\theta} + \frac{m\,B_{\mathrm{PM}}}{J} \theta = \frac{m\,B_{\mathrm{drive}}(t)}{J} \, ,
\end{eqnarray}
which is a differential equation for a driven harmonic oscillator with the resonance frequency:
\begin{eqnarray}
\label{torquesscalar1}
f_{\mathrm{res}}=\frac{1}{2\pi}\sqrt{\frac{m}{J}B_{\mathrm{PM}}}\,.
\end{eqnarray}
In this model, the oscillator damping (friction) has been neglected, which we find to be a good approximation for a compass. When the frequency $f_{\mathrm{drive}}$ of the drive field matches the resonance frequency of the compass needle, the needle experiences a resonant excitation, and the oscillation amplitude drastically increases. When the frequencies do not match, the drive results just in a slight jitter of the needle. 

In our experiment, we first fix the frequency $f_{\mathrm{drive}}$ of the current through the coil. Then we slide the permanent magnet closer to or farther from the compass and so tune the field $B_{\mathrm{PM}}$ experienced by the compass. The field of the permanent magnet in the dipole approximation depends on the distance $d$ between the centers of the magnet and the compass as $B_{\mathrm{PM}}=\frac{\mu_0}{4\pi}\frac{2m_{\mathrm{PM}}}{d^3}$, where $m_{\mathrm{PM}}$ is the magnetic moment of the permanent magnet. From Eq.\,\ref{torquesscalar1}, the resonance frequency of the needle is therefore 
\begin{eqnarray}
\label{fnat}
f_{\mathrm{res}}=\frac{1}{2\pi}\sqrt{\frac{\mu_0}{2\pi}\frac{m}{J} \frac{\,m_{\mathrm{PM}}}{d^{3}}}\,.
\end{eqnarray}

\begin{figure}[b]
\includegraphics[width = 0.5\textwidth]{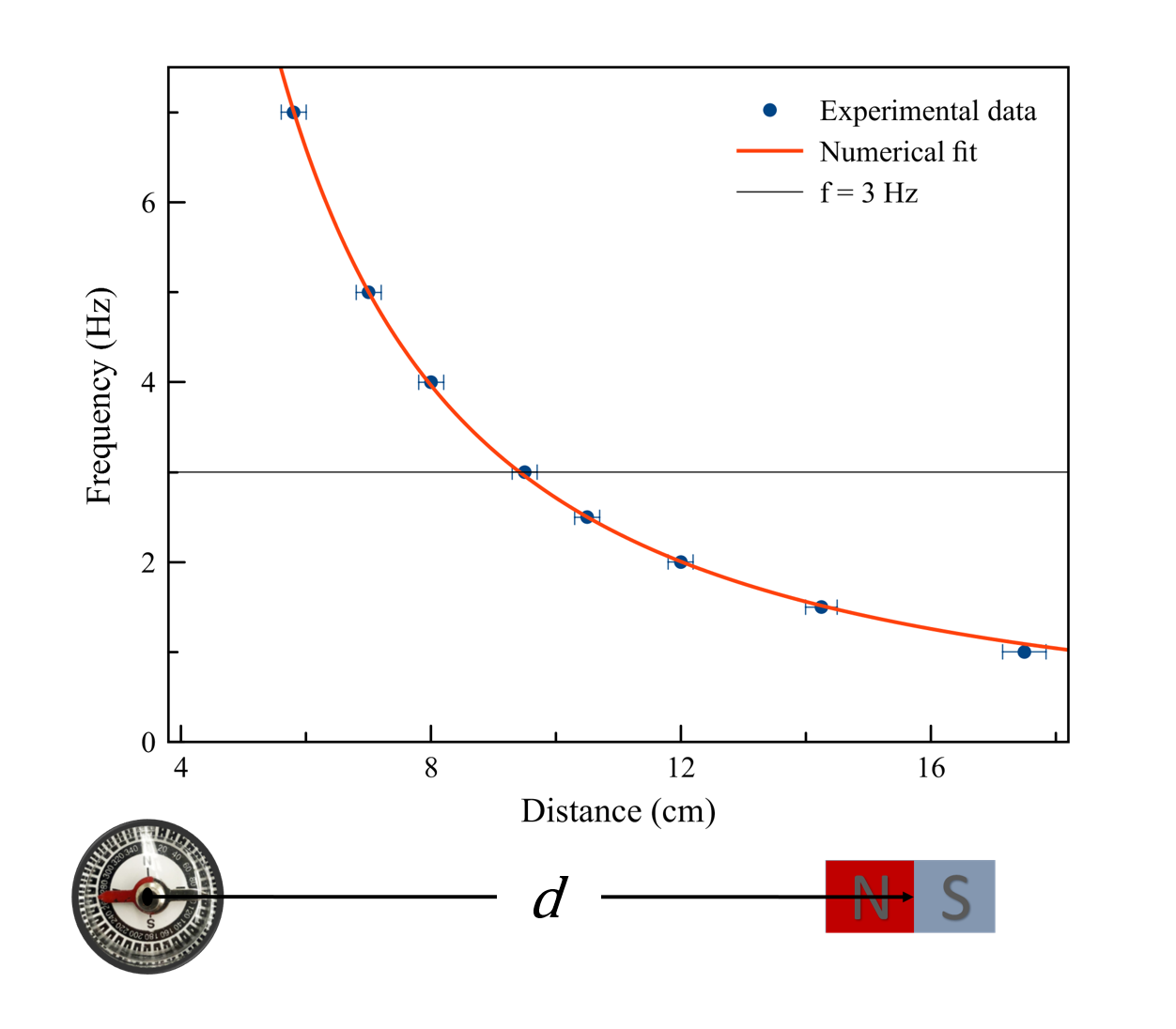} 
\caption{Resonance frequency as a function of distance $d$ between the compass and the  permanent magnet. Red curve is the fit using Eq.\,6.}
\end{figure}

By moving the permanent magnet along the $x$-axis, the resonance condition $f_{\mathrm{res}}(d)=f_{\mathrm{drive}}$ can be realized. Figure~2 shows the resonance frequency experimentally evaluated as a function of distance $d$ with the setup shown in Fig.\,1(d). The data is fitted well to Eq.\,\ref{fnat} (introducing a systematic error of 1\,cm to $d$).

\section{4. Experiment design and operation}

To be clearly visible, the frequency of the oscillator should be in the range of a few Hertz. Furthermore, the "sweet spot" distance of the permanent magnet should be in the manageable range of 5-15\,cm. A cheap thumbnail-sized neodymium permanent magnet possesses a magnetic moment of approximately $m_{\mathrm{PM}}\sim 250$\,Amp\`ere$\cdot$m$^2$. The magnetic moment of a compass needle depends on its size, so does the moment of inertia. The latter can be approximated to $J=\frac{1}{12} \rho_{\mathrm{needle}} V_{\mathrm{needle}} S^2$, where $\rho_{\mathrm{needle}}$ is the density of the needle material and $S$ is the size of the compass (length of the needle). Equation~\ref{fnat} for the resonance frequency can thus be rewritten as 
\begin{eqnarray}
\label{needle}
f_{\mathrm{res}}=\frac{1}{2\pi}\sqrt{\frac{\mu_0}{2\pi}\frac{12\,M_{\mathrm{needle}}}{\rho_{\mathrm{needle}} S^2} \frac{\,m_{\mathrm{PM}}}{d^{3}}}\, .
\end{eqnarray}

Equation \ref{needle} provides guidance to designing the experiment. It show that the resonance frequency is inversely proportional to the size of the compass. The resonance frequency also scales with the magnetic moment of the permanent magnet. Combining several magnets together, such as shown in Figs.\,1(c-d), can be used to increase the resonance frequency.

For a thumbnail-sized compass, we (empirically) estimate the magnetic moment to 0.86$\times10^{-3}$ \,Amp\`ere$\cdot$m$^2$ and the moment of inertia to $1.03 \times 10 ^{-11}\,$kg$\cdot$m$^2$. The resonance frequency of the compass at $d=9.5$\,cm should therefore be about 3\,Hz. As shown in Fig.\,2, this is indeed what we measure.

The size of the coil should be chosen such as to leave sufficient space for the compass and not to block the view. We use a coil with 6.35\,cm diameter, 1.25\,cm length, and 800 windings. Such coils are commercially available and inexpensive. A coil can  alternatively be fabricated by the educator ($\sim 100$ windings of any thin, flexible, insulated wire are preferable).


As a source for the alternating current, the audio output jack of a computer can be used. The coil wires are soldered to a 1/8'' inch stereo phone plug (old headphones can be repurposed \footnote{Exchange one speaker by the coil, check your audio device specifications}). With a free tone generator software or with an internet website \footnote{Several free websites and apps are available upon web search for "free tone generator"}, single frequency signal can be generated from the audio output of the computer. The volume level should be adjusted such that the compass needle oscillations in resonance do not exceed 30\,deg.

Finally, the compass and the permanent magnet need to be positioned on the plane intersecting the center of the coil. A supporting stage can be fabricated from any nonmagnetic material, e.g. cardboard. Alternatively, the stage can be 3D-printed using plastic. On request, the authors will provide a list of recommended components and the model file for 3D-printing.

\section{5. Compass versus spin resonance}
In science, magnetic resonance phenomena refer to resonant excitation of spins -- this can be either the spins of the atomic nuclei or the spins of the electrons. The spin of these microscopic particles is an inherent inalterable property and leads to an associated magnetic moment. In a magnetic field, the magnetic moments experience a torque that tends to align them parallel to the field. An additional alternating magnetic drive excites an oscillation of the moments. Leaving the quantum nature of the spin aside, the main difference of the compass resonance is that the needle oscillates within the compass plane. Magnetic moments of the particles, on the other hand, are bound to the inherent angular momentum -- the spin of the particles. Their motion is therefore precessional, similar to the motion of a spinning top. The precessional frequency is most generally given by:
\begin{eqnarray}
\label{fspin}
f_{\mathrm{spin}}=\frac{1}{2\pi}\gamma B\, , 
\end{eqnarray}
where $\gamma$ is the so-called gyromagnetic ratio -- the ratio of magnetic moment of the particle to its inherent angular momentum. The gyromagnetic ratio of atomic nuclei is smaller than that of the electrons by approximately 1000 (by virtue of the different masses), such that the typical frequencies for nuclear magnetic resonance lie in the range of MHz and for electron magnetic resonance -- in the range of GHz. The magnetic fields used in such experiments typically range from 10\,mT to several Tesla.

\section{6. Learning objectives}

The overall learning objective of this demonstration is to foster scientific literacy in science students. The presentation of the experiment is adaptable to student audiences of different preparedness levels and allows for incrementally increasing the depth of understanding of physical concepts.  

For high school students, the learning objectives may involve: describing the basics of electromagnetism in terms of fields,  combining theoretical concepts with a physical situation, and extending the concept of the restoring force in a harmonic oscillator to fields.

For university students, the learning objectives can be further extended to: predicting the frequency of oscillations, estimating the magnetic field of a permanent magnet and electromagnetic coil, estimating the magnetic moment of a magnetic object, recognizing when approximations are useful and when to use them,  converting a physical configuration to the junior level mathematical model.


\section{Acknowledgements}
This work was supported by the National Science Foundation under Grant No.~ECCS-1810541. D.L.M. acknowledges support by California Science Project. We thank Ward Beyermann for fruitful discussions.


\end{document}